\documentclass[aip,rsi,reprint,graphicx]{revtex4-1} 
\usepackage{natbib}
\usepackage{subfigure}
\usepackage{latexsym}
\usepackage{amssymb}
\usepackage{ulem}
\usepackage{float}
\usepackage{graphicx}
\usepackage{amsmath}
\usepackage{bm}
\usepackage{epsfig} 
\usepackage{multirow}
 \usepackage[bookmarks=false]{hyperref}

\begin{document}

\title{A millikelvin all-fiber cavity optomechanical apparatus for merging with ultra-cold atoms in a hybrid quantum system}
\author{H. Zhong}
\author{G. Fl\"{a}schner}
\author{A. Schwarz}
\email[]{Electronic address: aschwarz@physnet.uni-hamburg.de}
\author{R. Wiesendanger}
\affiliation{Institut f\"{u}r Angewandte Physik, Universit\"{a}t Hamburg, Jungiusstrasse 9-11, 20355 Hamburg, Germany}

\author{P. Christoph}
\author{T. Wagner}
\author{A. Bick}
\author{C. Staarmann}
\author{B. Abeln}
\author{K. Sengstock}
\author{C. Becker}
\email[]{Electronic address: cbecker@physnet.uni-hamburg.de}
\affiliation{ZOQ - Zentrum f\"{u}r Optische Quantentechnologien, Universit\"{a}t Hamburg, Luruper Chaussee 149, 22761 Hamburg, Germany}

\date{\today}

\begin{abstract}

We describe the construction of an apparatus designed to realize a hybrid quantum system comprised of a cryogenically cooled mechanical oscillator and ultra-cold $^{87}$Rb atoms coupled via light. The outstanding feature of our instrument is an \textit{in-situ} adjustable asymmetric all-fiber membrane-in-the-middle cavity located inside an ultra-high vacuum dilution refrigerator based cryostat. We show that Bose-Einstein condensates of $N=2\times10^6$ atoms can be produced in less than 20\,s and demonstrate a single photon optomechanical coupling strength of $g_0=2\pi\times9$\,kHz employing a high-stress Si$_3$N$_4$ membrane with a mechanical quality factor $Q_{\rm m}>10^7$ at a cavity set-up temperature of $T_{\rm MiM}=480$\,mK.

\end{abstract}


\maketitle

\section{Introduction}

\begin{figure*}[ht!]
\centering
\includegraphics[width=13.5cm]{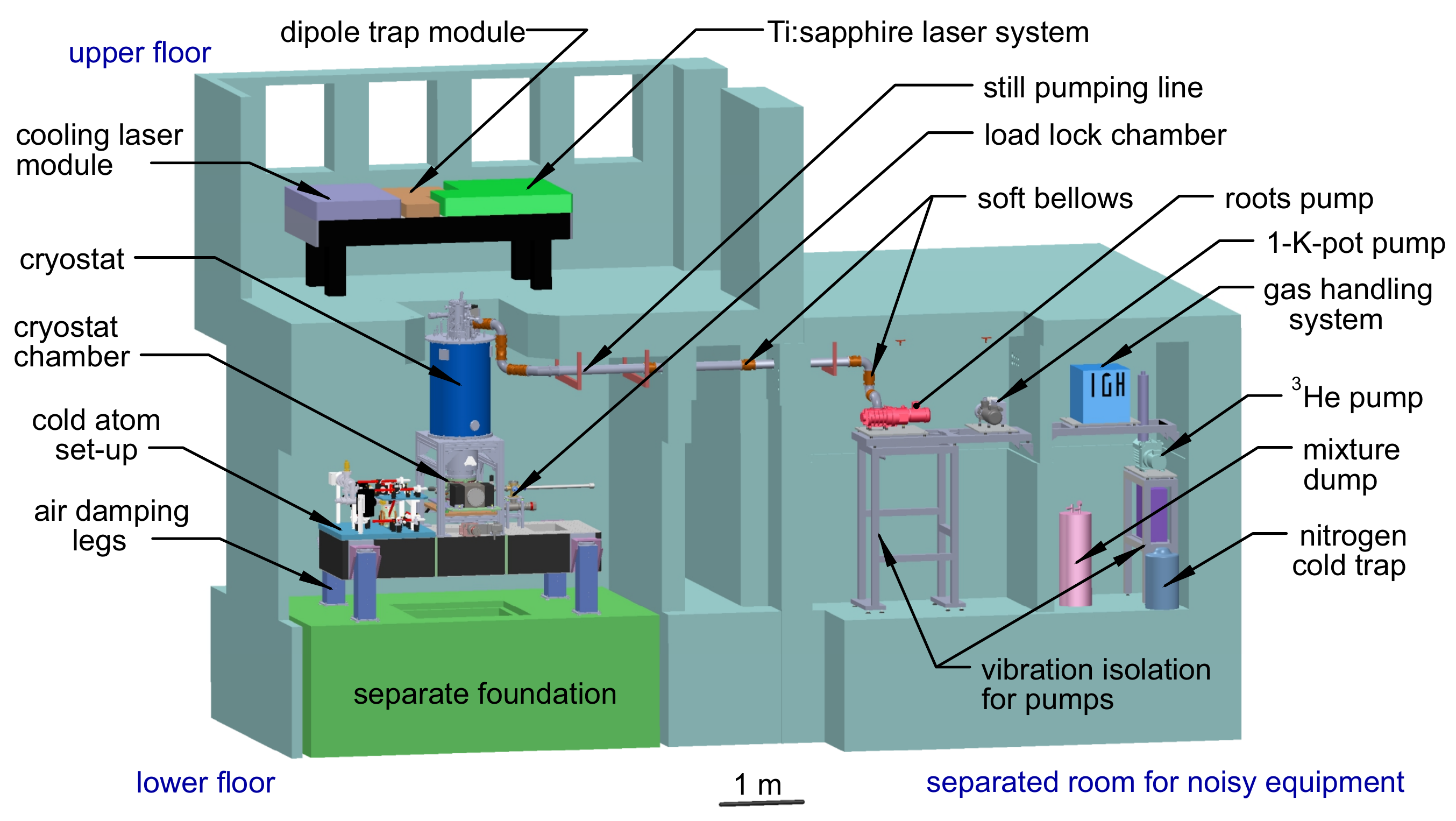}
\caption{Overview of the laboratory with the experimental set-up designed to realize a membrane-atom HQS. The UHV cryostat with the all-fiber MiM cavity inside (Sec.\,\ref{cuhv}), the cold atom set-up (Sec.\,\ref{bec}), and the homodyne detection scheme (Sec.\,\ref{cls}) are placed on an optical table in the lower floor, which stands on a separate foundation. The coupling and detection laser system (Sec.\,\ref{cls}) is assembled on a second  optical table that is located in the upper floor, together with the experimental control and data acquisition system. All mechanical pumps used to run the DR as well as its gas handling system, the $^3$He/$^4$He-mixture dump and the nitrogen cold trap are kept in a separate side room.}
\label{Abb:lab}
\end{figure*}

A hybrid quantum system (HQS), \citep*{Kurizki2015} consists of two or more different physical quantum objects ideally combining complementary functionality in terms of e.g., manipulation, storage and detection of quantum states.
The advent of quantum optomechanics \citep*{Aspelmeyer2014} with the possibility to cool mechanical oscillators to their quantum ground state, \citep*{Connell2010, Teufel2011, Chan2011} has opened up new perspectives to use them as one integral part of such HQS owing to the inherent ability of mechanical motion being able to `interact with everything'. 
Once realized, such devices can be utilized for highly sensitive optical detection of small forces and displacements, for manipulation and detection of mechanical motion in the quantum regime or as a coherent light-matter interface for quantum information processing \citep*{Aspelmeyer2014} to name only a few examples.
Another particularly interesting class of objects for a HQS are atomic systems such as trapped ions or cold atoms,\citep*{Vogell2013, AJ2015} or artificial atoms like, e.g., spins, nitrogen vacancy (NV) centers in diamond or superconducting qubits.\citep*{Xiang2013} 
These systems are characterized by conceptual similarities in preparation and read-out of quantum states, but are distinct in view of other important properties like their coupling strength to other quantum systems or their isolation from environmental influences. Particularly, atomic systems possess very long coherence times compared to other quantum systems due to their weak interaction with the environment. 
Hence, several proposals have been put forward for coupling cold atoms with mechanical oscillators,\citep*{Vogell2013, Hammerer2009, Wallquist2010, Bariani2014, Bariani2015}, e.g., the interaction of the magnetic moment of a Bose-Einstein condensate (BEC) with a nanomagnet on a cantilever,\citep*{Treutlein2007} or coupling a vibrating mirror to atoms in an optical lattice, which can drive the transition between a Mott insulator and a superfluid state.\citep*{Chen2009}
Few experimental realizations of combining nanomechanical oscillators with cold atoms have been reported.\citep*{AJ2015, Wang2006, Hunger2010} However, none of these experiments is operated with both subsystems in their quantum ground states.
Up to date the quantum ground state of a mechanical oscillator has been either reached by cryogenically cooling a GHz resonator \citep*{Connell2010}, or by resolved sideband cooling of integral nanomechanical structures in optomechanical systems, both in the microwave \citep*{Teufel2011} and optical \citep*{Chan2011} domain.
The realization and detection of entanglement between microscopic and macroscopic degrees of freedom in such a HQS is of major interest, \citep*{Hammerer2009, Wallquist2010} either for studying the quantum to classical transition or for quantum information applications.

In this paper we report on the construction and characterization of an experiment designed to realize a HQS comprised of ultracold $^{87}$Rb atoms and a cryogenically cooled mechanical oscillator coupled to each other via light.
Our experiment is designed to meet the specific requirements of a strongly coupled HQS. In particular, the optical finesse of our optomechanical set-up strongly deviates from the optimal value for a pure optomechanical experiment. To reach the strong coupling regime the total decoherence rate has to be smaller than the coupling strength. \citep*{Vogell2013}  We envisage experiments ranging from sympathetic cooling of the membrane by cold atoms, \citep*{Vogell2013, AJ2015, Bariani2014} to entangling the membrane motion with an internal atomic degree of freedom. \citep*{Hammerer2009, Vogell2015} For all these experiments it is beneficial to have a membrane phonon occupation number as low as possible.
We designed and built two dedicated set-ups, each individually optimized to house one part of the HQS. Ultracold atomic samples of $^{87}$Rb are produced in an ultra-high vacuum apparatus with excellent optical access, whereas the mechanical oscillator, a Si$_3$N$_4$ membrane, is enclosed in an asymmetric planar-concave all-fiber cavity, \citep*{Bick2016} which is attached to a $^3$He/$^4$He dilution refrigerator (DR). This so-called `membrane-in-the-middle' (MiM) configuration, \citep*{Thompson2008} allows optimizing its optical and mechanical properties independently and can be currently cooled down to temperatures $T_{\rm MiM} <500$\,mK in a UHV environment with pressures $p < 1\times10^{-9}$\,mbar.
We decided for a UHV compatible DR to allow for direct \textit{in-situ} atom-oscillator coupling inside the cryostat in a future experimental stage. \citep*{Jessen2014} 

Resonant coupling of the motional degrees of freedom of cold atoms trapped in an optical lattice to the motion of a Si$_3$N$_4$ membrane oscillator practically limits the frequency of the employed resonator to $f<400$\,kHz.\citep*{Vogell2013, AJ2015}
For such low-frequency mechanical oscillators cryogenic cooling to the quantum ground state is not possible. To accomplish this, further optomechancial cooling schemes have to be applied.
The primary goal of a HQS is the achievement of strong coupling.
That means that the coupling has to be larger than the total decoherence rate in the combined system. Main contributions to that rate include resonant light scattering at the atoms, cavity loss and membrane heating through light absorption and light pressure shot noise.
A thorough analysis reveals that an optimal regime for strong atom-membrane coupling exists for a cavity finesse of $F\approx 500$. \citep*{Vogell2013} Taking into account the very short cavity length of a mode-matched all-fiber cavity ($L_{\rm cav} \approx 25$\,$\mu$m) \citep*{Bick2016} and the corresponding very large free spectral range (FSR $\approx 6$\,THz), the typical linewidth of such a medium finesse all-fiber cavity is very broad ($\Delta \nu \approx 12$\,GHz). 
Thus, we are well outside the so-called `resolved sideband regime' and the corresponding optomechanical cooling schemes \citep*{Teufel2011, Chan2011} cannot be applied efficiently for our system. 
However, ground state cooling of a low-frequency mechanical oscillator in the sense that the mean phonon occupation number is smaller than one, $n_{\rm m} \ll 1$, can be achieved by velocity-dependent active feedback cooling. \citep*{Poot2012} This is however only possible if the environmental temperature is low enough and the detection noise is reduced down to standard quantum limit. \citep*{Poot2012} The temperature requirement is met in our experiment by placing the MiM cavity set-up in a DR. 

The paper is organized as follows: Section\,\ref{lab} gives a brief overview of our laboratory layout. The two most important components of our instrument are the optomechanical cavity set-up inside a DR and the ultra-cold $^{87}$Rb atoms apparatus discussed in Sec.\,\ref{cuhv} and Sec.\,\ref{bec}, respectively. The coupling laser system is discussed in Sec.\,\ref{cls}.
Finally, Sec.\ref{test} provides characterization measurements that demonstrate the feasibility to realize a HQS with our set-up that is based on ultra-cold atoms and a cryogenically precooled low frequency mechanical oscillator.

\section{Laboratory layout}
\label{lab}

The overall layout of the two-storey laboratory for the HQS experiment is depicted in Fig.\,\ref{Abb:lab}. 
The upper floor contains the coupling laser module, the dipole trap module and the titanium sapphire (Ti:sapphire) laser module, which are assembled on an optical table. The optical table is supported by air damping legs and a temperature controlled flow box on top ensures stable conditions. The control units as well as the data acquisition systems are located in the upper floor as well.

The UHV cryostat system and the ultra-cold atom apparatus including the homodyne detection are mounted on a second optical table in the lower floor. This table is also supported by air damping legs and for even better vibration isolation rests on a separate foundation. Above the area with the optical components another temperature controlled flow box is installed. A hole in the ceiling provides access to the upper floor, from where the cryostat is supplied with liquid nitrogen and liquid helium. 

Equipment that generates mechanical vibrations like the pumps to operate the DR and the chiller unit to cool the magnetic traps are located in a side room. All pumps are mounted on rubber feet and soft bellows are used along the still and 1-K-pot pumping lines to reduce transmission of vibrations to the cryostat. The gas handling system, the cold traps to remove contaminations from the $^{3}$He/$^{4}$He-mixture and the dump to store the mixture, if the DR is not in operation, are situated in the side room as well. 

\section{Cryogenic UHV system}\label{cuhv}

The cryogenic UHV system consists of a small load lock chamber to introduce mechanical oscillators into the system and a larger chamber with the cryostat on top. 
With our design membranes can be exchanged \textit{in-situ}, i.e., without warming-up and without breaking the ultrahigh vacuum. 
The DR-insert with the cavity set-up is loaded from above into the cryostat so that it extends into the UHV chamber below. This design allows mechanical access to exchange membranes \textit{in-situ} as well as optical access to align the all-fiber cavity \textit{in-situ}. 

As pointed out above, cryogenically precooling a low-frequency mechanical oscillator is essential to reduce its initial mean phonon occupation number. We decided for a DR cryostat, because simpler cryogenic cooling methods cannot achieve a base temperature in the millikelvin regime while providing a large cooling power.
The latter is important to compensate for optical absorption in the membrane and heating due to stray light. Additionally, the UHV environment in the cryostat minimizes damping losses due to the interaction with and adsorption of residual gas molecules. 

\subsection{Dilution refrigerator}\label{cryo}
    
\begin{figure}
\centering
\includegraphics[width=8.5cm]{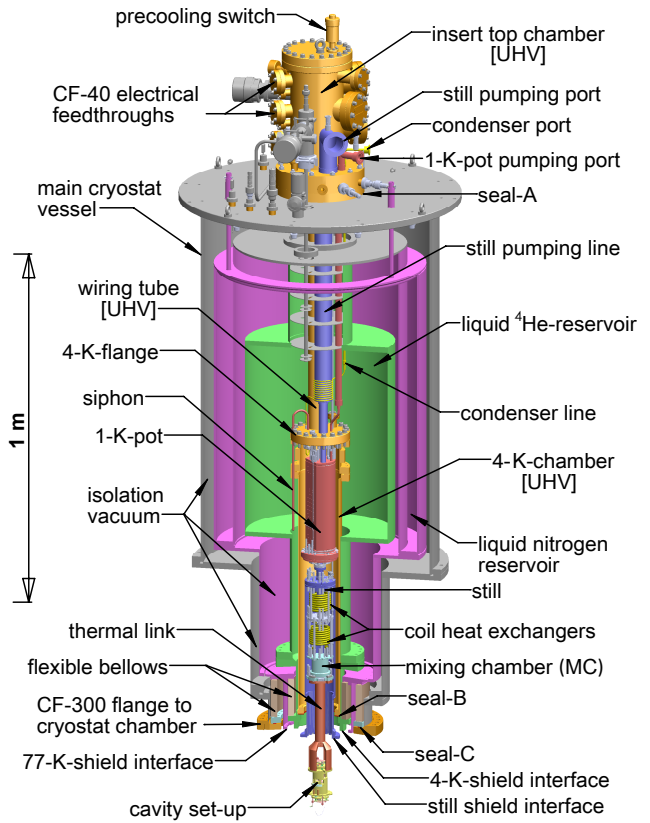}
\caption{Section view of the cryostat. The main cryostat vessel encloses the top-loaded insert, the liquid nitrogen reservoir (pink) and the liquid $^{4}$He-reservoir (green). The isolation vacuum ($p< 10^{-4}$\,mbar) is maintained between the two cryogen reservoirs and between the liquid nitrogen reservoir and the outer cryostat vessel. On the insert, the major UHV components (bright yellow, marked by `UHV') include top chamber, wiring tube and 4-K-chamber. The DR unit (including 1-K-pot, still, two stainless steel coil heat exchangers and the MC) is topped by the 4-K-flange of the 4-K-chamber immersed inside the $^{4}$He-reservoir. Details of three seals: (A) a Viton seal separating the $^{4}$He-reservoir from air, (B) a non-standard CF interface (with fasteners in the inward cycle of the knife edge) separating the tail of the $^{4}$He-reservoir from the UHV area, and (C) a CF-250 interface (also with fasteners inwards) separating the isolation vacuum from the UHV area.} 
\label{Abb:cryo} 
\end{figure}

A detailed description of cryogenic cooling schemes, cryostat design, thermal anchoring and thermometry can be found in, e.g., Ref.\,\onlinecite{Pobell}. In the following we restrict ourselves to the specifics of our DR cryostat. It is based on a modified version of the Kelvinox 400HA from Oxford Instruments\citep*{OI} and shown in Fig.\,\ref{Abb:cryo}. Full UHV-compatibility, meaning that the DR can be baked by internal heaters up to $100^\circ$C, is achieved by using CF-type flanges and VCR connectors from Swagelok\citep*{Swagelok} instead of indium seals. Furthermore, we work without step heat exchangers, which deteriorate if heated to $100^\circ$C, and instead employ two coil heat exchangers. The main cryostat vessel with the insulation vacuum houses the UHV insert that is surrounded by an inner liquid $^{4}$He-reservoir with a volume of 82\,l (6 days hold time), and an outer liquid nitrogen reservoir with a volume of 65\,l (4 days hold time).

The UHV insert consists of a still pumping line and a condenser line, a 1-K-pot pumping line, a top chamber with all electrical feedthroughs and a wiring tube through which all electrical wires are guided into the 4-K-chamber with the DR unit inside.
After attaching the cavity set-up via the thermal link to the MC, the insert is loaded from above into the main cryostat vessel, and connected at Seal-A and Seal-B, as shown in Fig.\ref{Abb:cryo}. Subsequently, the still-, 4-K-, and 77-K-shields (cf. Sec.\ref{shield}) are attached to their corresponding interfaces located at the bottom of the cryostat. Finally, the fully assembled cryostat is loaded onto the cryostat chamber (cf. Sec.\ref{chamber}). In this way, the UHV-conditions are maintained inside the UHV components of the insert (cf. Fig.\ref{Abb:cryo}) and the cryostat chamber (cf. Fig.\ref{Abb:uhv}). Two (edge-welded) flexible bellows at the bottom of the cryostat are introduced to avoid structural damage due to different thermal contraction upon cool down and also provide long thermal paths between different temperature stages for thermal decoupling. Additionally, they also separate the isolating vacuum from the UHV area via Seal-C near the bottom of the cryostat. 

To replenish the 1-K-pot with $^{4}$He a siphon is fed through the 4-K-flange into the  $^{4}$He reservoir. The $^{4}$He-flow can be adjusted by a needle valve. If the 1-K-pot is full and the needle valve closed (single shot operation) the hold time is more than 12\,h. In this mode of operation vibrational noise due to intermixing of incoming normal $^{4}$He with the superfluid $^{4}$He inside the 1-K-pot can be avoided.\citep*{Gorla2004} 

All electrical wires for the heaters and temperature sensors in the DR unit as well as for the piezo elements used to drive the stepper motors of our two fiber alignment stage in our cavity set-up (cf. Sec. \ref{mim}) enter the UHV system via electrical feedthroughs at the insert top chamber. They are successively thermally anchored with Vacseal \citep*{SPI} at the 4-K-flange, the 1-K-pot, the still, the cold plate between the two heat exchangers and finally the MC. To minimize thermal conduction via the electrical connection manganin and constantan wires with diameters of 100\,$\mu$m are used to connect the piezo elements and the temperature sensors. Commercially available $\pi$ filters (60\,dB attenuation at 100\,MHz) are directly mounted on-top of the electrical feedthroughs outside the UHV to reduce radio frequency interference and heating. 

To measure the temperatures at the different stages of the DR unit, several temperature sensors are installed: a ruthenium oxide (ROX) sensor at the bottom of the MC, a ROX temperature sensor at the cold plate between two continuous heat exchangers, a ROX sensor at the still, and a ROX and a Cernox sensor at top and bottom of the 1-K-pot, respectively. Carefully calibrated ROX sensors can measure temperatures down to a few mK. To calibrate the ROX sensor at the MC, we used a $^{60}$Co nuclear orientation thermometer. \citep*{Pobell} Figures of merit for the performance of a DR are its base temperature and its cooling power at 100\,mK. Without the cavity set-up attached to the MC we measured a base temperature of $T_{\rm MC}= 31.2 \pm 0.2$\,mK. Using a heater next to the MC to regulate $T_{\rm MC}$ to 100\,mK, we determined a cooling power of 560\,$\mu$W.

A cylindrical thermal link made from high conductive oxygen free copper (OFHC) is used between MC and cavity set-up to bring its final position to the center of the cryostat chamber (see Fig.\,\ref{Abb:cryo}). At this position it is possible to monitor the alignment of the fibers and to exchange membrane \textit{in-situ}. Coin silver (90\% Ag with 10\% Cu) screws are used, because their thermal contraction is larger than that of copper at low temperatures, which ensures that the pressed contacts between MC and thermal link as well as between cavity set-up and thermal link becomes tighter after cooling down the system. 
	
Initial cool down to the base temperature in the millikelvin regime takes about 4.5 days. To accelerate the cooling process of the thermally well isolated DR unit, a lozenge-shaped copper piece is used as a mechanical heat switch.\citep*{Kingsley2012} It is operated by a CF-16 UHV rotary feedthrough attached to the top chamber of the insert with its driving rod extending down to the MC. By rotating the thermal switch a mechanical contact can be established or released between the bottom of the 1-K-shield and a strike-plate mounted on the MC.
				
\subsection{Radiation shields with movable shutters}\label{shield}

\begin{figure}
\centering
\includegraphics[width=7.3cm]{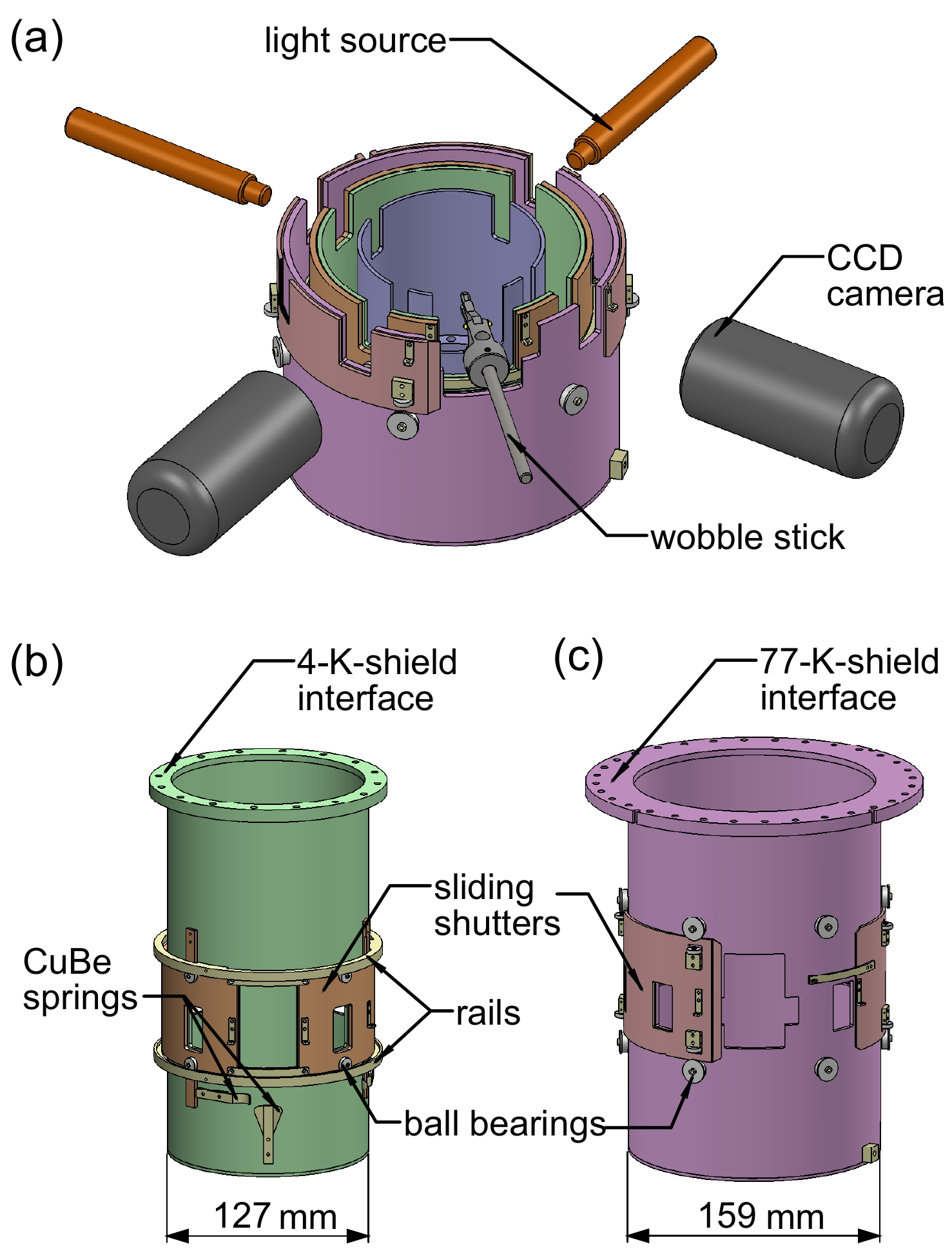}
\caption{Home-built UHV compatible radiation shields with apertures for optical and mechanical access. \textbf{(a)} Perspective view of three concentrically arranged radiation shields including the positions of the two CCD cameras and the corresponding light sources for illumination. The wobble stick used to \textit{in-situ} exchange membranes is also shown. \textbf{(b)} and \textbf{(c)} show the fully assembled 4-K-shield and 77-K-shield, respectively. Both feature stainless steel ball-bearing supported rotating shutters that can be moved using the wobble stick.} 
\label{Abb:shield} 
\end{figure}

To avoid heating by thermal radiation, home-built cylindrical metallic shields shown in Fig.\,\ref{Abb:shield}(a) surround the cavity set-up.
They are connected to the liquid nitrogen reservoir (77 K), the $^{4}$He-reservoir (4.2 K) and the still (0.7 K), respectively.
In order to monitor \textit{in-situ} the alignment of the fiber cavity at low temperatures, four symmetrically placed rectangular apertures are made in each shield. They are in-line with four CF-63 viewports on the cryostat chamber (Sec.\,\ref{chamber}). Two orthogonally placed CCD cameras with diffuse light sources at opposite windows are used to observe the positions of the two fibers relative to the membrane inside the cavity. To access the membrane shuttle receptacle with a wobble stick, an additional rectangular aperture is made on all three shields. As shown in Fig.\,\ref{Abb:shield}(b) and (c), the five apertures in the 4-K-shield as well as in the 77-K-shield can be opened and closed with a rotating shutter system. We do not use shutters for the apertures in the still shield, because tests with apertures in the still shield but no apertures in the 4-K-shield and 77-K-shield have shown no significant increase of $T_{\rm MC}$.

To move the rotating shutters handles attached to them can be grabbed with the wobble stick. Three positions can be selected: all apertures closed (measurement condition), all apertures open except the one for the wobble stick (during aligning the fiber position relative to the membrane), and all apertures open (exchanging membranes). All the shields and the shutters are made from oxygen free highly conductive copper (OFHC; UNS C10100) and phosphorous deoxidized copper (DHP-Cu; UNS 12200). To minimize the emissivity and to avoid surface degradation due to oxidation over time the copper shields are plated with about 5\,$\mu$m of gold. Flexible copper braids are used to connect the rotating shutters to the radiation shields. Smooth motion of the shutters is ensured by using UHV-compatible oil-free stainless steel ball bearings. 
 
Before the shutters are opened the $^{3}$He/$^{4}$He-mixture has to be put back into the storage dump, because proper circulation cannot be maintained in the presence of this large additional thermal load from room temperature radiation through the open apertures. As a result, the temperature increases continuously over time to about 30\,K, if the apertures are open to align the cavity or to exchange the membrane. Subsequent cool down back to the base temperature requires a few hours, depending on the amount of time the apertures were open.
	
\subsection{Ultra-high vacuum chamber assembly}\label{chamber}
		
\begin{figure}
\centering
\includegraphics[width=8.3cm]{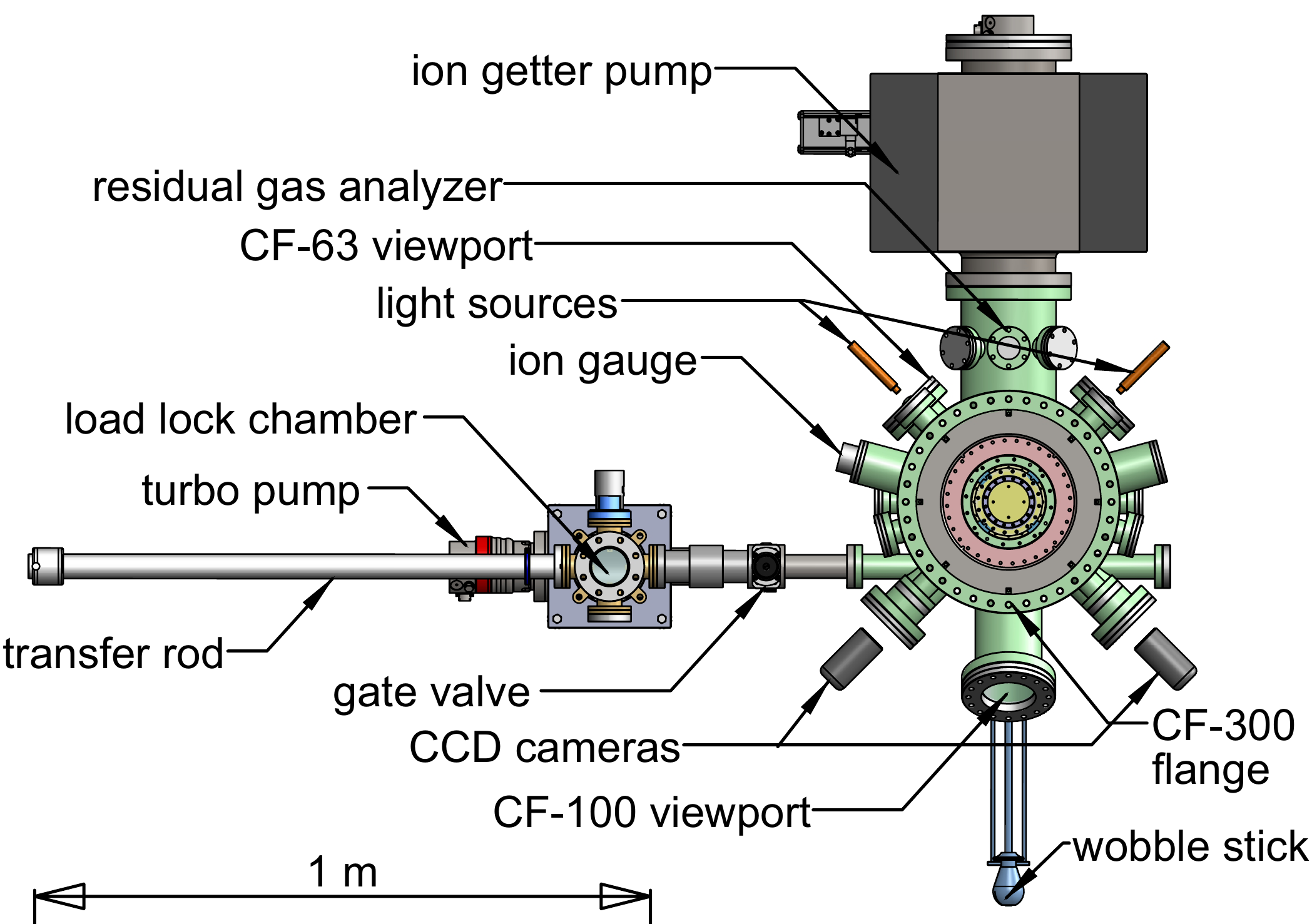}
\caption{Top view of the UHV chamber assembly. The CF-300 flange of the cryostat chamber is connected to the bottom of the cryostat vessel. A load lock chamber, equipped with a magnetically driven transfer rod, is attached to the cryostat chamber via a Viton sealed CF-40 gate valve. The wobble stick, as well as the two cameras and the two light sources used to align the cavity are also shown.} 
\label{Abb:uhv} 
\end{figure}

Figure\,\ref{Abb:uhv} shows the UHV chamber assembly. The cryostat chamber is made of non-magnetic stainless steel (main vessel: 316LN; tubes welded to the vessel: 316L). After manufacturing, a heat treatment (vacuum firing and subsequent bake-out at 200\,$^{\circ}$C for 48\,h) has been performed to remove hydrogen dissolved inside the stainless steel material, which reduces hydrogen outgassing. Each chamber is pumped by a turbo molecular pump backed by an oil-free scroll pump. During experiments, these pumps are switched off and the cryostat chamber is only pumped by a vibrationless 300\,l/s ion getter pump. The pressure in the UHV system is monitored with an ionization gauge. To detect potential air or helium leaks, a residual gas analyzer is attached to the cryostat chamber.
 
The cryostat chamber is equipped with four CF-63 viewports, in line with the four sets of optical apertures in the radiation shields (Fig.\, \ref{Abb:shield}). As described in the previous section, they are used for optical access. A wobble stick with a jaw style pincer head \citep*{Ferrovac} in line with the fifth set of apertures in the radiation shields can be used to exchange membranes and to open or close the rotating shutters. All actions with the wobble stick can be observed through an angled CF-100 viewport above the wobble stick flange. 

The whole UHV system including the insert can be baked up to 100\,$^{\circ}$C. The maximum temperature is limited by the materials used to build the DR unit. However, pressures below $\leq$ 5$\times$10$^{-10}$\,mbar are achieved even without baking if the system is cold, because the large surface area of the radiation shields acts as a powerful cryopump.
	
\subsection{All-fiber MiM cavity set-up}\label{mim}

The mode-matched all-fiber MiM cavity set-up is the central component of the optomechanical part of our experiment. It is an asymmetric cavity with different reflectivities of the dielectric coatings at the fiber ends, respectively.\citep*{Bick2016} Both fibers enter the UHV system via CF-16 Swagelok fiber feedthroughs with Teflon gaskets, \citep*{Abraham1998} which are attached to the bottom of the UHV-cryostat chamber. They are guided through tiny holes (1 mm diameter) in the radiation shield assembly and glued into zirconia (ZrO$_{2}$) ferrules. Each ferrule is glued to the free end of a piezo tube (PZT-8, outer diameter $d= 3.2$\,mm, thickness $t = 0.38$\,mm and length $l=20$\,mm), used to fine-adjust or to scan the cavity length.

\begin{figure}
\centering
\includegraphics[width=8cm]{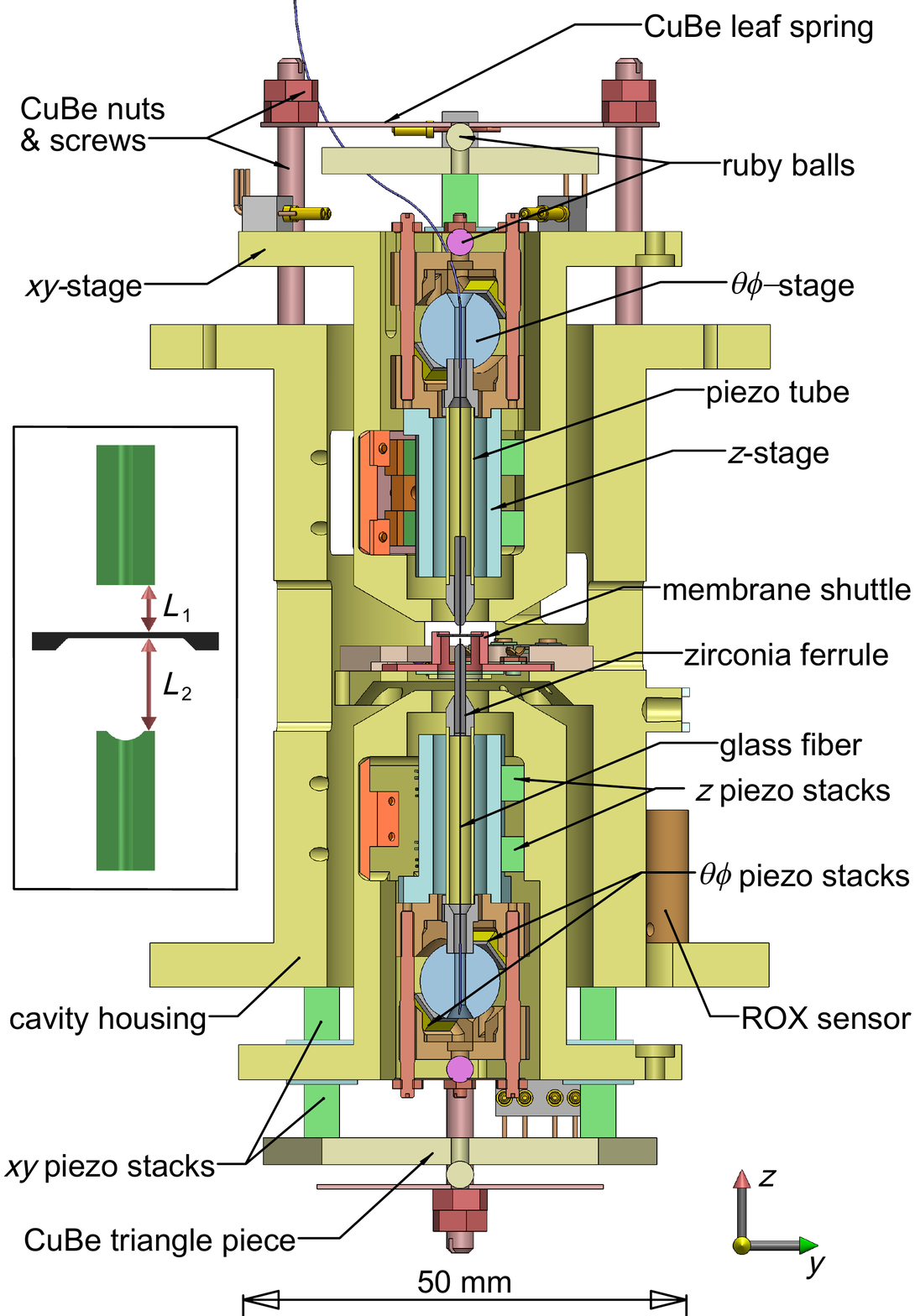}
\caption{Section view of the all-fiber MiM cavity set-up. The magnified area in the inset shows the two fibers and the membrane with its frame in between. The cavity housing encloses two identical piezo-driven positioning stages with a $z$-, $xy$-, and $\theta\phi$-stepper motor, which provide five degrees of freedom for each fiber. The membrane receptacle (cf. Fig. \ref{Abb:shuttle}) is located between them. Both fibers are glued into a piezo tube to fine adjust their $z$-positions and to scan the cavity length $L_{\rm cav}$ as well as the fiber-membrane spacings $L_1$ and $L_2$ (see inset). The MiM cavity temperature $T_{\rm MiM}$ is measured with a ROX sensor as well as with a Cernox sensor. Both are attached next to each other to the cavity housing.} 
\label{Abb:mim} 
\end{figure}

To align the MiM cavity, each fiber can be fully adjusted independently with five degrees of freedom ($x,y,z,\theta$, and $\phi$) by piezo-driven Pan-type slip-stick stepper motors \citep*{PanPatent}. Translational and angular accuracy are better than 1 nm and better than $0.1^\circ$, respectively. Thus, it is possible to ensure that the fibers are in-line and perpendicular to the membrane. Figure\,\ref{Abb:mim} shows the section view of our cavity set-up with the two alignment stages and the membrane in between. It has an outer diameter of 70\,mm with an overall height of 135\,mm. The cavity housing is made from gold-plated (5$\mu$m) OFHC copper, which ensures a fast cool down and a homogeneous temperature distribution. The gold plating avoids surface degradation (during assembling and maintenance the cavity set-up in air) and enhances the thermal conductivity through pressed contacts. Most of the screws and all leaf springs are made from CuBe (UNS C17200). Two types of epoxy glues are used, i.e., H20E (electrically conductive) and H77 (electrically insulating), respectively.\citep*{Epotek} All piezo stacks used for the stepper motors have been purchased from PI.\citep*{PI} More technical details regarding the construction of such a positioning stage, how to apply voltages to the piezo stacks to induce the desired motion and how to adjust the step size have been described in Ref.\,\onlinecite{PanPatent} for single-axis $z$-stage, in Ref.\,\onlinecite{Liebmann2002} for dual-axis $xy$-stage and in Ref.\,\onlinecite{Zhong2014} for dual-axis $\theta\phi$-stage. 

\begin{figure}
\centering
\includegraphics[width=7.8cm]{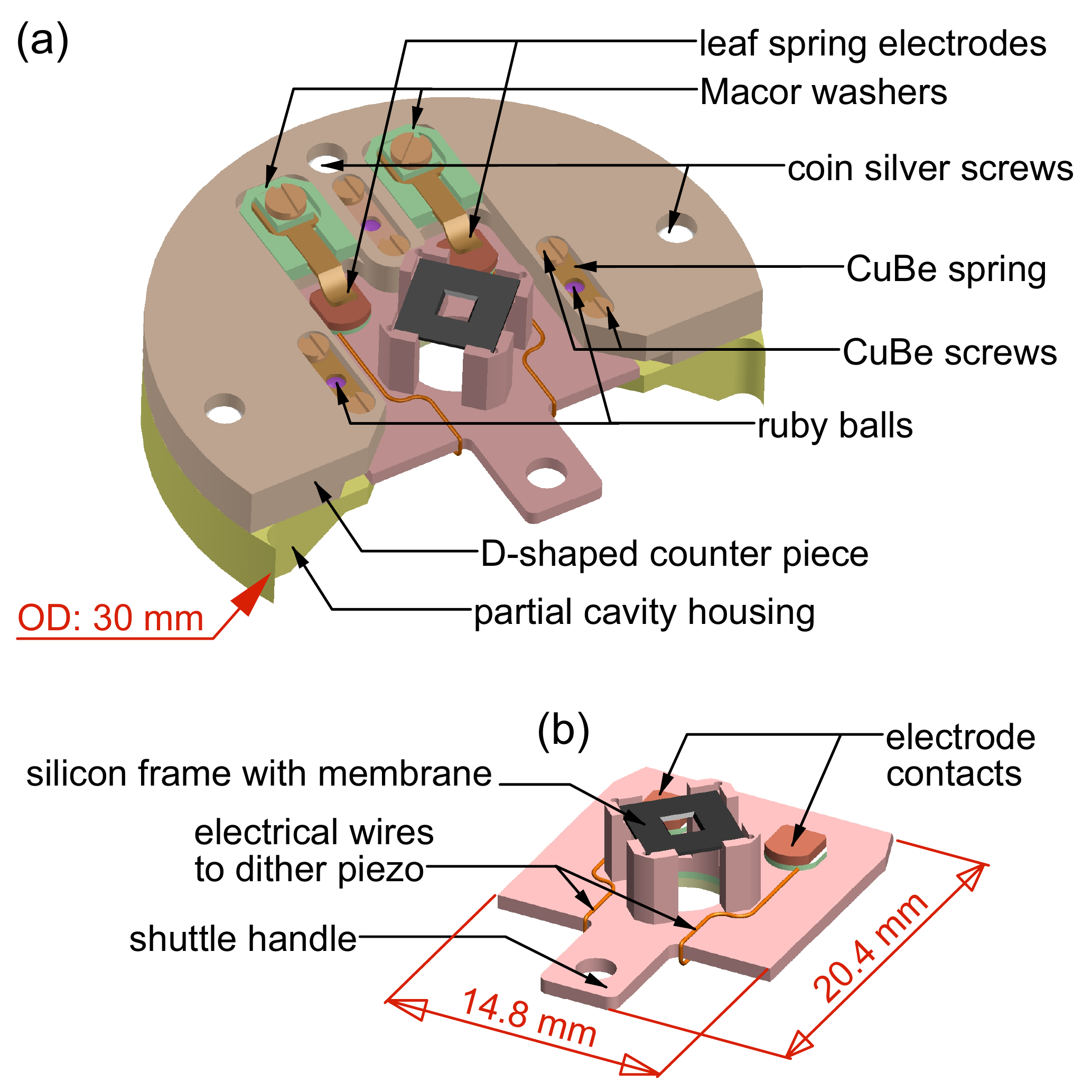}
\caption{\textbf{(a)} The shuttle receptacle and \textbf{(b)} the shuttle carrying a membrane. A dither piezo can be glued to the backside of the shuttle and excited via the electrode contacts on the shuttle and the leaf spring electrodes in the receptacle.} 
\label{Abb:shuttle} 
\end{figure}

Figure\,\ref{Abb:shuttle}(a) shows the shuttle receptacle, which is located in the middle of the cavity housing between the two fiber ends, cf. Fig.\,\ref{Abb:mim}. The receptacle is a 1\,mm slit formed between an integral part of the cavity housing and a D-shaped counter piece attached to it with coin silver screws. If the shuttle is inserted into the slit, three spring loaded ruby balls clamp it tightly. Additionally, two electrical contacts, isolated from the metallic surroundings, are formed via two CuBe leaf springs, which strike two electrically isolated contact pads on the shuttle if inserted. They can be used to drive a dither piezo attached to the shuttle to excite the membrane.
Note that after inserting the shuttle, the position of the membrane is fixed and not adjustable. Further alignment of the MiM cavity is done by independently moving each fiber end with respect to the membrane surface in $x$, $y$, $z$, $\theta$ and $\phi$ directions.    

The shuttle together with a Si$_{3}$N$_{4}$ membrane in its frame is depicted in Fig.\,\ref{Abb:shuttle}(b). The shuttle is a rectangular CuBe plate with a central 5 mm hole surrounded by four posts, onto which the silicon frame of the membrane is glued. A handle with a hole, which can be grabbed with the wobble stick, sticks out to one side. Before inserting or extracting the shuttle from the receptacle, both fibers must be fully retracted, so that they cannot be touched by the wobble stick or the shuttle. During the cavity alignment, one fiber is approached from top and the other from below towards the membrane. While the upper fiber is always visible with the CCD cameras, the lower fiber has to approach through the hole in the shuttle. Thereafter, its position can be seen through the gaps between the four posts. Upon further approach towards the membrane it becomes invisible again, because of the 0.2\,mm thick silicon frame (see inset in Fig.\,\ref{Abb:mim}). Hence, to align the lateral positions of the fiber cores relative to each other and relative to the membrane with a precision of about 10 nm, the cavity signal in transmission and reflection has to be monitored (see also Sec.\,\ref{test}).
To determine the total cavity length $L_{\rm cav}$ with a precision better than 1 nm, successive cavity resonances generated with the Ti:sapphire laser described in Sec.\,\ref{cls} can be utilized. Moreover, $L_1$ and $L_2$ can be determined with an accuracy of about 100\,nm using white light interferometry.
Our method used here is very similar to the one described in Ref.\onlinecite{Jiang2008} for determining the length of very short Fabry-P\'{e}rot cavities.

\section{Cold atom set-up}\label{bec}

\begin{figure*}
\centering
\includegraphics[width=11.5cm]{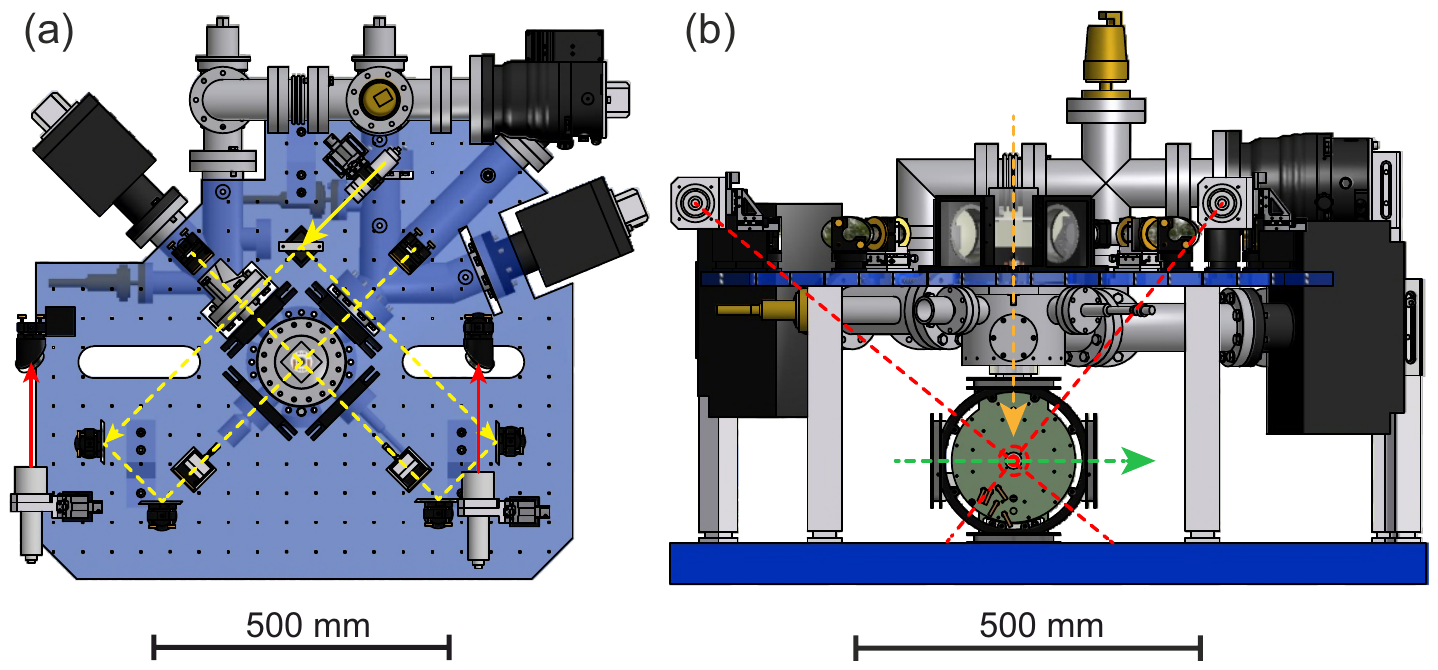}
\caption{Drawing of the experimental apparatus to produce a BEC. \textbf{(a)} Top view showing upper breadboard and 2D-MOT setup. \textbf{(b)} Side view showing upper and lower glass cells as well as magnetic trap (green housing) and magnetic field compensation coil cube. Also shown are the beam paths for the 2D-MOT (yellow), 3D-MOT (red), detection beam (green), and pushing beam (orange). The 2D-MOT beam is split on the upper breadboard using a polarizing beam splitter and expanded employing a cylindrical telescope to form elliptic beams as required for the 2D-MOT. The upper 3D-MOT telescopes are located on the upper breadboard. The corresponding beams are reflected through elliptical holes in the breadboard onto the lower glass cell. Both beams are rotated by $8^\circ$ with respect to the $45^\circ$ axes which is reserved for a retro-reflected 2D lattice (not shown here). The pushing beam passes vertically through the differential pumping tube in the center of the vacuum chamber and transfers the atoms from the 2D-MOT into the 3D-MOT.} 
\label{Abb:mots} 
\end{figure*}

To produce ultra-cold $^{87}$Rb atoms a dedicated apparatus has been constructed and placed next to the cryostat on the same optical table in the lower floor; cf. Fig.\,\ref{Abb:lab}. The set-up (Fig.\,\ref{Abb:mots}) is based on a scheme consisting of a two-dimensional magneto- optical trap (2D-MOT) to catch atoms from a background gas and a 3D-MOT operating at pressures below $1 \times 10^{-11}$\,mbar.
The two different vacuum glass cells are connected via a differential pumping stage allowing for pressures that can differ by a factor of $10^3$.
This setup has the advantage of providing extremely good optical access to the lower 3D-MOT glass cell allowing for different kinds of optical trapping, manipulation and detection schemes, such as optical lattices of different dimensionality, Raman laser configurations or momentum resolved Bragg spectroscopy. \citep*{Ernst2010}

In a typical experimental sequence we start by loading the 3D MOT for $10\,$s resulting in atom numbers of $N=10^{10}$ at temperatures of $T \approx T_{\mathrm D}$, where $T_{\mathrm D}= 146\,\mu$K is the Doppler temperature of $^{87}$Rb.
Subsequently, the atoms can be further cooled in an optical molasses reducing the temperature to $T_{\mathrm{min}} = 10\,\mu$K, which amounts to several times the $^{87}$Rb recoil temperature of $T_{\mathrm{rec}}=362\,$nK.
For the purpose of generating a Bose-Einstein condensate (BEC), we load our atoms in a magnetic trap of hybrid cloverleaf 4D type.
Forced evaporation cooling for less than $20\,$s allows producing Bose-Einstein condensates of $N_{\mathrm{BEC}} \approx 2\times 10^6$ particles without any discernible amount of thermal atoms.
For continuous experiments our setup is equipped with a crossed optical dipole trap derived from a Nd:YAG laser operated at $1064\, $nm with circular beam waists of $52\,\mu$m and $242\,\mu$m. The maximum available optical power at the experiment is $8\,$W per beam.
The BEC inside this dipole trap has an elongated cigar like shape and corresponding trapping frequencies $(\omega_y,\omega_z) = 2 \pi \times (85, 62) $\,Hz, where gravity points along the $z$-direction.
The beam with the larger beam waist can be used to tune $\omega_x$ between $2\pi \times (1 \ldots 20)$\,Hz. 
Thereby we can vary the elongation of the atomic cloud along the direction of the coupling lattice or coupling Raman beams, respectively.
For experiments aiming at coupling internal atomic degrees of freedom to the motional state of a mechanical oscillator it is necessary to trap the atoms in a potential that is independent of the particular internal state to avoid fast dephasing. This is guaranteed by using a far detuned optical dipole potential. \citep*{Grimm2000}
From the same laser we also derive a two-dimensional optical lattice perpendicular to the coupling lattice (see Sec. \ref{cls}) that enables us to confine the atoms in a three-dimensional periodic potential and to freeze out all continuous degrees of freedom.
The properties of the cold atom sample are detected using a flexible absorption imaging system allowing for different magnification ranging from 0.5 to 10.

\section{Coupling laser system} \label{cls}

\begin{figure*}
\centering
\includegraphics[width=13cm]{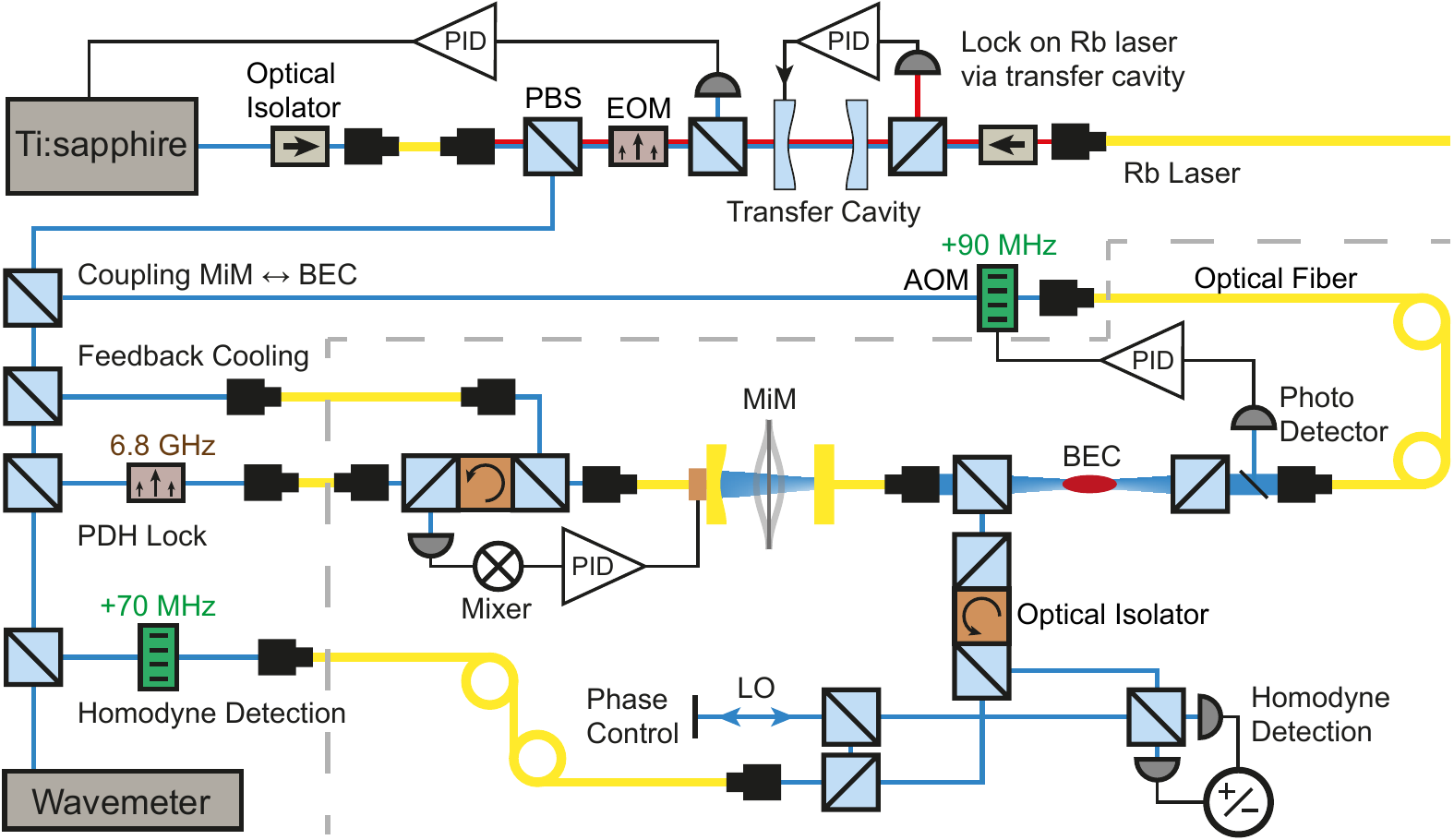}
\caption{Coupling and detection laser system based on a titanium-sapphire (Ti:sapphire) laser locked via a transfer cavity to the $^{87}$Rb cooling laser system. Four different beam branches are derived by polarizing beamsplitter cubes and are guided to the lower floor laboratory using polarization maintaining fibers (yellow lines). The coupling beam is frequency shifted and intensity controlled by an acousto-optic modulator (AOM), passes the atomic cloud (indicated by the red ellipse), and is then coupled into the cavity via the planar fiber. The homodyne detection beam is superimposed with the coupling lattice beam and also coupled to the MiM cavity from the planar side owing to the much better mode match. LO denotes the local oscillator beam of the homodyne detection, and PBS denotes polarizing beam splitter. The Pound-Drever-Hall (PDH) locking beam is phase modulated with an electro-optical modulator (EOM) and superimposed with the feedback cooling beam using an optical isolator and coupled into the cavity via the curved fiber end.}
\label{Abb:laser}
\end{figure*}

Figure\,\ref{Abb:laser} shows the layout of the coupling laser system. To achieve a coupling between the membrane and the ultra-cold atomic cloud, we use a titanium-sapphire laser (Ti:sapphire) operated at a wavelength close to the $^{87}$Rb D2 line at a wavelength of $\lambda = 780\,$nm. This laser can be widely tuned and Pound-Drever-Hall-locked to an external Fabry-P\'{e}rot transfer cavity with a free spectral range (FSR) of 1GHz, which is referenced to the $^{87}$Rb cooling laser system. The light is divided into four different branches that serve as the coupling beam, the homodyne detection beam to determine the membrane motion, the Pound-Drever-Hall-locking beam for the fiber Fabry-P\'{e}rot cavity of the MiM set-up and the feedback beam for active feedback cooling of the membrane. The coupling and detection beams are coupled into the MiM setup from the planar side to allow optimal mode match and thus minimized losses.\citep*{Bick2016} For the less critical feedback and Pound-Drever-Hall beams we choose an incoupling from the curved side of the cavity. Two AOMs in the coupling and detection beam, respectively, shift the frequencies by several tens of MHz in order to prevent unwanted interference between the four different beams. Furthermore, these AOMs can be used to control and actively stabilize the light power in these beams. The coupling beam is focused onto the atomic cloud with a waist size of $w_0 = 78\,\mu$m and then coupled into the MiM set-up. Afterwards, the reflected light from the MiM set-up interferes with the incoming beam and forms a 1D optical lattice at the position of the atoms providing the basis of the coupling scheme. Our detection beam has a typical power of $1-5\,\mu$W at the MiM system and the emerging phase modulation due to the motion of the membrane is resolved through interference with a phase-locked local oscillator beam in a balanced homodyne detection scheme. In order to lock the cavity length of the MiM set-up to the Ti:sapphire laser wavelength we use a Pound-Drever-Hall technique employing sidebands at 6.72\,GHz created by a high-frequency resonant phase modulator. The back-reflected beam is detected by a fast photodiode, amplified by two low noise amplifiers with a total gain of 90\,dB and mixed down with the local oscillator that drives the phase modulator.
In order to modulate the light intensity for the active feedback cooling we use a fiber based, large bandwidth amplitude modulator, which is driven by a digital oscillator. This signal is derived digitally from the homodyne signal after processing through a lock-in amplifier. \citep*{ZI} 

\section{Characterization Measurements}\label{test}

With the cavity set-up attached to the MC and with the rotating shutter system attached to the radiation shields but without in-coupling laser light a minimal temperature $T_{\rm MiM}= 473$\,mK is currently achieved. If the roots pump is switched off, i.e., only a rotary vane pump is used to circulate the $^{3}$He/$^{4}$He-mixture, the temperature increases to 478\,mK. Since this temperature increase is negligible, all measurements shown in this section were performed without roots pump. With a homodyne detection power of 5 $\mu$W $T_{\rm MiM}$ increases further to 485\,mK. At the same time the temperature at the MC is $T_{\rm MC}= 178$\,mK. Since the cooling power of the non-loaded MC is about 560\,$\mu$W at $T_{\rm MC}= 100$\,mK, the total thermal load can be estimated to be $\dot{q} \approx 1$\,mW, if we assume a linear relationship between cooling power and temperature from 100\,mK to 178\,mK.\citep*{Pobell}. 

We believe that heat radiation is responsible for most of the total thermal load. We estimate an upper bound for the thermal load caused by thermal conduction along the electrical wires of less than 1\,$\mu$W and along the two fibers of less than 11\,$\mu$W and find it to be negligible.  
Since we do not use shutters on the still-shield, the cavity set-up is directly exposed to radiation from the 4-K-shield. However, we estimate the heat load due to direct radiation heating to be less than 10\,$\mu$W, even if the actual temperature of the 4-K-shield is about 10\,K. Another path for heat radiation is funneling: The shutters of the 4-K-shield are about 4\,mm larger in size than the apertures they close. Since a gap of about 1\,mm exists between the 4-K-shield and the shutters, heat radiation from the 77-K-shield can reach the cavity set-up via multiple reflections between the gold plated inner surface of the shutter and the gold plated outer surface of the 4-K-shield. The amount of heat load by funneling is difficult to estimate. However, after we moved the shutters in front of the 4-K-shield sideways by about 1\,mm, so that there is still no direct line of sight between 77-K-shield and cavity set-up, $T_{\rm MiM}$ increased by about 10\,mK. This finding indicates that currently the shutters do not prevent funneling completely. Possible solutions would be to increase the shutter size relative to the aperture size or to implement shutters on the still-shield.

To analyze the feasibility to realize a HQS with our experimental set-up, we characterize the performance of our millikelvin optomechanical set-up with a high stress Si$_{3}$N$_{4}$ membrane (nominal dimensions $a \times a \times t = 1.5\,{\rm mm}\times 1.5\,{\rm mm} \times50\,{\rm nm}$) from Norcada \citep*{Norcada}. The asymmetric mode-matched cavity employed for the following measurements consists of a planar fiber end with a nominal reflectivity $|r_{1}|^{2}=0.907 \pm 0.004$ (for incoupling) and a curved fiber end (radius of curvature $r_{\rm ROC}\approx 50$~$\mu$m) with a nominal reflectivity $|r_{2}|^{2}=0.995 \pm 0.001$ (for outcoupling).
	
\subsection{Fiber cavity alignment}\label{alignment}
	
\begin{figure}
\centering
\includegraphics[width=8.5cm]{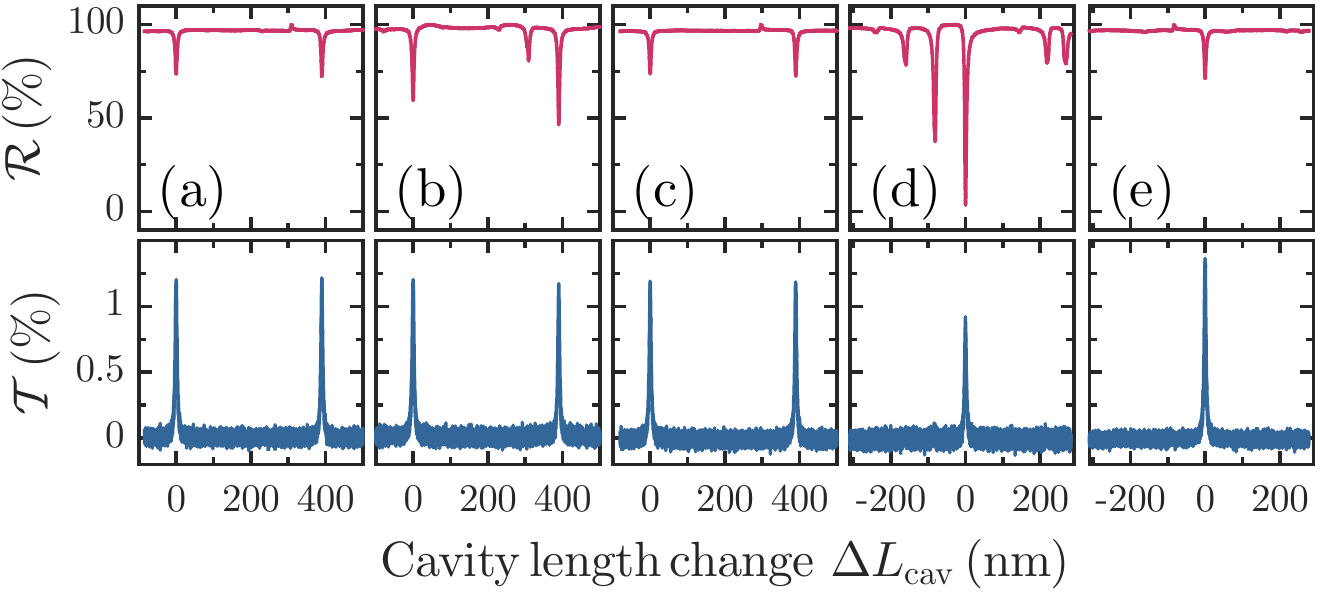}
\caption{Repeatability test of the coarse \textit{z} motion by measuring the empty cavity power transmission $\mathcal{T}$ and power reflection $\mathcal{R}$ at $T_{\rm MiM} \approx 25$\,K as a function of cavity length change $\Delta L_{\rm cav}$. \textbf{(a)} After optimal alignment of the fibers with an empty cavity length of $L_{\rm cav}\approx 15\,\mu$m. One fiber retracted by 0.4\,mm, and then reapproached (same $L_{\rm cav}$): \textbf{(b)} without any further lateral or angular alignment, and \textbf{(c)} with slight \textit{xy} realignment. Both fibers retracted into their $xy$-housings and then reapproached (same $L_{\rm cav}$): \textbf{(d)} without any further lateral or angular alignment, and \textbf{(e)} with slight \textit{xy} realignment. The zeros of the horizontal axes in (a)-(e) are given with respect to their respective resonance positions. The calibration factor to convert the horizontal axis to nm is deduced from the free spectral range being $\lambda/2 = 390$\,nm in our case.}  
\label{Abb:fc} 
\end{figure}		
		
In our set-up it is very important that the cavity can be aligned reliably at low temperatures. We analyze the repeatability of the fiber cavity alignment using the empty fiber cavity. The alignment as well as the measurements shown in Fig.\,\ref{Abb:fc} were performed at $T_{\rm MiM} \approx 25$\,K with the inspection apertures open and the $^3$He/$^4$He-mixture in the dump. The on-resonance cavity exhibits its designed 75\% intensity reflectivity and 1.5\% intensity transmission, which are very close to the calculated values using the nominal reflectivities of the fiber coatings.

From Fig.\,\ref{Abb:fc}(a), obtained after thorough alignment of the cavity, we determined an empty cavity finesse $F_{\rm 0}\approx 60$, which is found to be unchanged up to a cavity length of 25\,$\mu$m.  In order to analyze the low-temperature alignment reproducibility we retracted one fiber by 400\,$\mu$m and then reapproached it to the original position. After that procedure the power reflectivity on resonance is slightly lower and additional features appear in the spectrum (see Fig.\,\ref{Abb:fc}(b)) indicating a decreased mode match. Comparison with independent measurements of the power reflectivity on resonance as a function of lateral fiber position suggests a lateral misalignment of $\Delta r_{\perp} < 1$\,$\mu$m. The original signal quality could be recovered using a few steps with the $xy$-stepper motors shown in Fig.\ref{Abb:fc}(c). The described procedure also works when retracting both fibers into their $xy$-housings ($\sim 8$\,mm apart) as required for exchanging the membrane shuttle, and the corresponding results are shown in Figs.\,\ref{Abb:fc}(d) and (e). The estimated lateral misalignment in Fig.\ref{Abb:fc}(d) is $\Delta r_{\perp} \sim 2$\,$\mu$m. 

\subsection{Homodyne detection of the membrane eigenmodes}
	
\begin{figure}
\centering
\includegraphics[width=8.5cm]{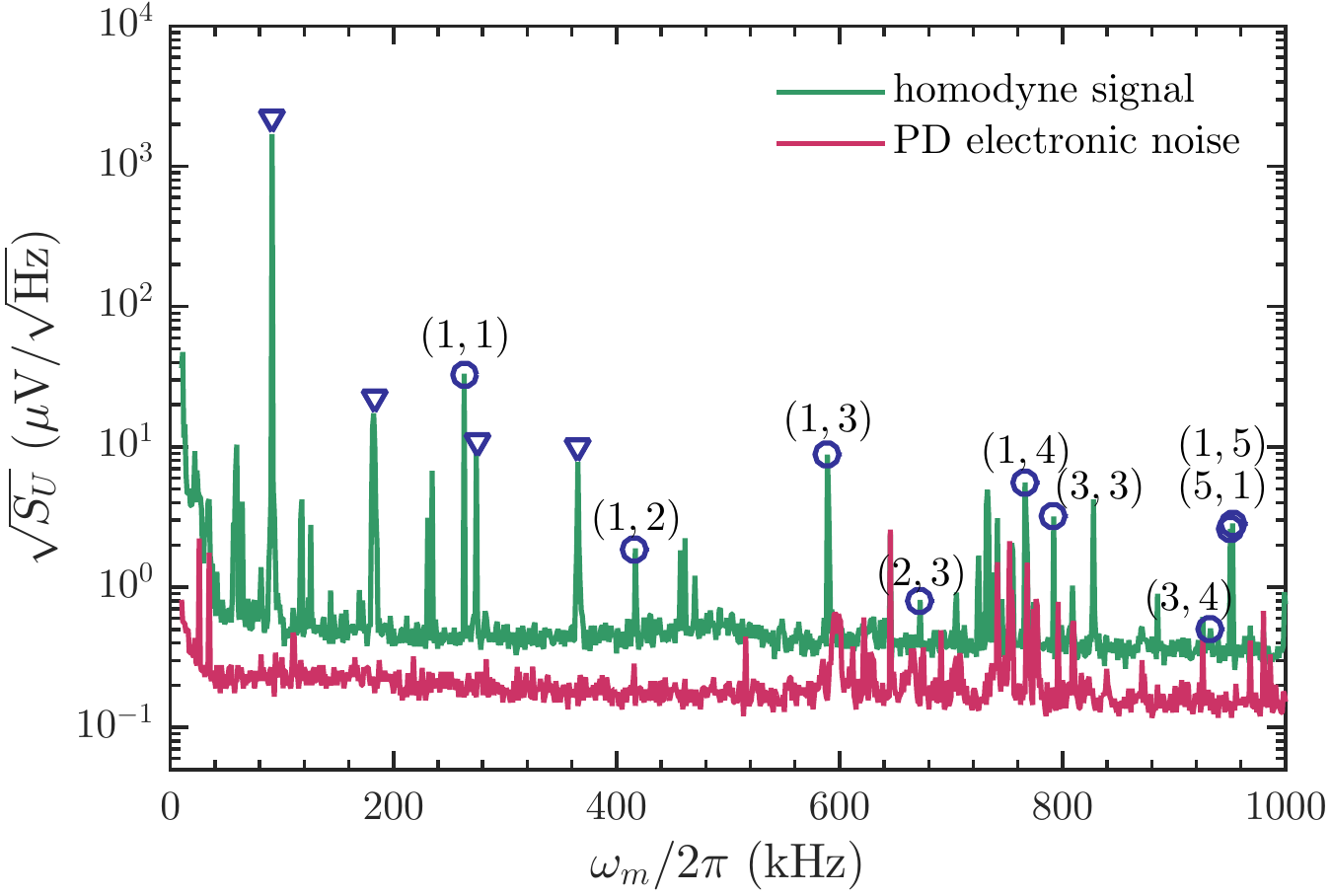}
\caption{Measurement of the membrane eigenmodes at $T_{\rm MiM}=480$\,mK (green) with a light power of 5\,$\mu$W. The spectrum was measured using our homodyne detection system with a noise floor of $\sqrt{S_{x_{\rm n}}} = 0.30 \pm 0.05$\,fm/$\sqrt{\rm Hz}$. Circles with numbers in brackets indicate the (m,n) eigenmode of the membrane. Triangles mark peaks stemming from the intrinsic 91.18 kHz modulation of the Ti:sapphire laser. The noise floor of the photo detector (PD, red) is shown for comparison.} 
\label{Abb:noise} 
\end{figure}

To identify the membrane eigenmodes and to characterize the different mechanical and electrical noise sources, we performed homodyne measurements of the thermal Brownian motion of a membrane over a span of 1\,MHz using a spectrum analyzer (Rohde \& Schwarz FSP)\citep*{SA} with a resolution bandwidth (RBW) of 10\,Hz, employing a laser power of 5\,$\mu$W at $T_{\rm MiM}=480$\,mK, as shown in Fig.\,\ref{Abb:noise}.
The homodyne detection was calibrated by simulating a defined membrane motion by moving both fiber tips commensurately with respect to the membrane. We used the signal generated by this well-defined membrane motion to determine the corresponding voltage per membrane displacement. 

We extracted a noise floor of $0.30 \pm 0.05$\,fm/$\sqrt{\rm Hz}$ around 270\,kHz from these measurements, which is on the same order of magnitude as the laser shot noise equivalent of 0.13 fm/$\sqrt{\rm Hz}$ for our set-up at 5 $\mu$W detection power. 
We further identified the different membrane eigenmodes $f_{m,n}$ by applying the formula $f_{m,n}=f_{1,1}\sqrt{(m^{2}+n^{2})/2}$ with $f_{1,1}=263.8$\,kHz. Since the measured amplitude of a given mode depends on the position of the almost point-like intracavity light mode (waist size $w_0=3$\,$\mu$m) with respect to the membrane mode function, monitoring a given eigenmode can be employed to optimize the coupling to a desired eigenmode by adjusting the lateral position properly. In Fig.\,\ref{Abb:noise} the (1,1)-eigenmode has been optimized by adjusting the two fiber positions with respect to the membrane center. For eigenmodes with a node at the membrane center (both $m$ and $n$ are even),  its amplitude becomes minimal. Consequently, (2,2)- and (2,4)-eigenmodes are not visible. For an ideal square membrane $f_{m,n}$ is equal to $f_{n,m}$, while the observed lifting of this degeneracy (cf., the splitting between the (1,5)- and (5,1)-eigenmodes with a frequency difference of $|f_{1,5}-f_{5,1}|\approx 2$\,kHz), indicates that our particular membrane is not perfectly square.

\subsection{Mode temperature and $Q$ factor}

\begin{figure}
\centering
\includegraphics[width=8.5cm]{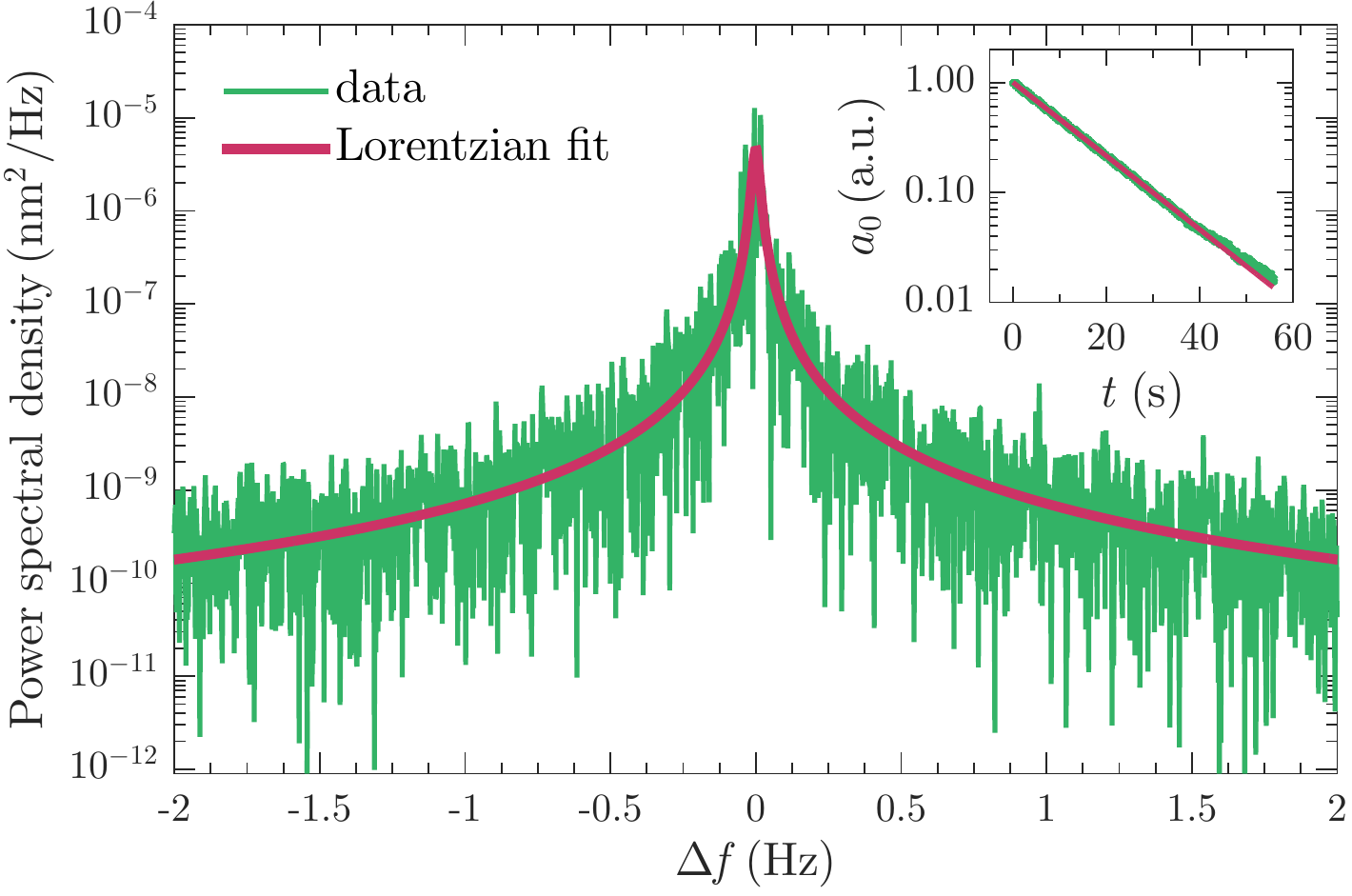}
\caption{Measured spectral density (single measurement) of the (1,1)-eigenmode of a high stress Si$_{3}$N$_{4}$ membrane at $T_{\rm MiM}=485$\,mK (green) fitted by a Lorentzian function (red). The noise spectrum was obtained using the zoom FFT function of a lock-in amplifier \citep*{ZI} with maximal 32768 FFT samples at a sampling rate of 56\,Hz (equivalent to the FFT span) for a data acquisition time of about 10\,min. To show the quality of the Lorentzian fit, only the central $\pm 2$\,Hz part of the recorded 56\,Hz full span curve with a frequency spacing of 1.7\,mHz is plotted. Inset: Logarithmic plot of a single ringdown measurement (green) recorded at $T_{\rm MiM}=480$\,mK, and fitted by an exponential decay function (red) in the form of $e^{-t/\tau_{\mathrm m}}$.}
\label{Abb:rd} 
\end{figure}

For the realization of an atom-membrane HQS the mode temperature $T_{1,1}$ (related to the phonon number $n_{\rm m}$) and the corresponding mechanical quality factor $Q_{\rm m}$ (related to the thermal decoherence rate $\gamma_{\rm m}n_{\rm m}\approx k_{\rm B}T_{1,1}/{\hbar}Q_{\rm m}$) of the membrane (1,1)-eigenmode are of key importance. The temperature of the (1,1)-eigenmode can be determined using the equipartition theorem according to $T_{1,1} = k\left\langle x^{2}\right\rangle/k_{\rm B}$, where $k=(2\pi f_{1,1})^2 \cdot m_{\rm eff}$ is the spring constant.
The effective mass is calculated using $m_{\rm eff}=\frac{1}{4} \rho a^2 t=9.7\times 10^{-11}$\,kg, \citep*{AJ2015} where $\rho_{\rm Si_3N_4}=3.44$ g/${\rm cm}^3$ is the density of Si$_{3}$N$_{4}$ membrane. \citep*{Borkje2012}
$\left\langle x^{2}\right\rangle=(1/2\pi)\times \int^{\infty}_{0} S_{x}(\omega)d\omega$ is obtained by integrating the fitted Lorentzian curve shown in Fig.\,\ref{Abb:rd}. 
For the Lorentzian fit, the experimentally determined noise floor of 0.3\,fm/$\sqrt{\rm Hz}$ is used as a fixed background and the linewidth $\gamma_{\rm m}$ is fixed to the value deduced from the ringdown measurement shown in the inset of Fig.\ref{Abb:rd}.
We believe that the quality of our fit indicates that the dephasing \citep*{Schneider2014} is not significant in our measurements, however, it cannot be fully ruled out without further studies in the future.
From the Lorentzian fit we obtain a mode temperature $T_{1,1} = 3.66 \pm 0.02$\,K, which is larger than the environmental temperature $T_{\rm MiM}=485$\,mK.
Reasons for the increased mode temperature could be radiative heating through the not perfectly overlapping rotating shutters, \citep*{Chan2011} temperature gradients between the membrane and the MiM set-up body,\citep*{DLMcAuslan2016} residual mechanical vibrations in the set-up, \citep*{Yuan2015} or electromagnetic noise picked up and transduced by the dither piezo. We have furthermore carefully checked that the mode temperature does not depend on the laser power used for homodyne detection, by changing the power by half an order of magnitude (up to 25\,$\mu$W) without observing any mode temperature change. 

The inset of Fig.\ref{Abb:rd} shows the normalized ringdown curve of the (1,1)-eigenmode obtained at a cavity set-up temperature of $T_{\rm MiM}=480$\,mK. Fitting an exponential decay function to the data we find a decay time $\tau_{\rm m}=(12.982\pm0.002)$\,s, which corresponds to $Q_{\rm m} = \pi f_{1,1} \tau_{\rm m} = 10,757,900\pm 1,500$. Similar large $Q$-factors for high stress Si$_{3}$N$_{4}$ membranes in this temperature regime have been reported recently.\citep*{Jayich2012,Yuan2015} The resulting mechanical decay rate deduced from the ringdown measurement reads $\gamma_{\rm m}=\omega_{\rm m}/Q_{\rm m}=2\pi\times 0.024$\,Hz.

\subsection{Optomechanical coupling}

Since our envisaged HQS relies on coupling ultra-cold atoms to a membrane via light, the strength of the optomechanical coupling is a crucial parameter. For a Fabry-P\'{e}rot cavity (length: $L_{\rm cav}$) with a moving end mirror, the optomechanical coupling strength is given by $G=\omega_{\rm cav}/L_{\rm cav}$, where the cavity frequency $\omega_{\rm cav}$ equals to the laser frequency $\omega_{\rm L}$. In the MiM configuration the optomechanical coupling strength is defined by $g_{m} = -\partial \omega_{\rm cav}/\partial z_{m}$, where $z_{m}$ is the intracavity position of the membrane. Using the optical methods mentioned in Sec.\,\ref{mim}, we find $L_{\rm cav}\approx23.7$\,$\mu$m, $L_1\approx5.8$\,$\mu$m and $L_2\approx17.9$\,$\mu$m.

\begin{figure}
\centering
\includegraphics[width=8.5cm]{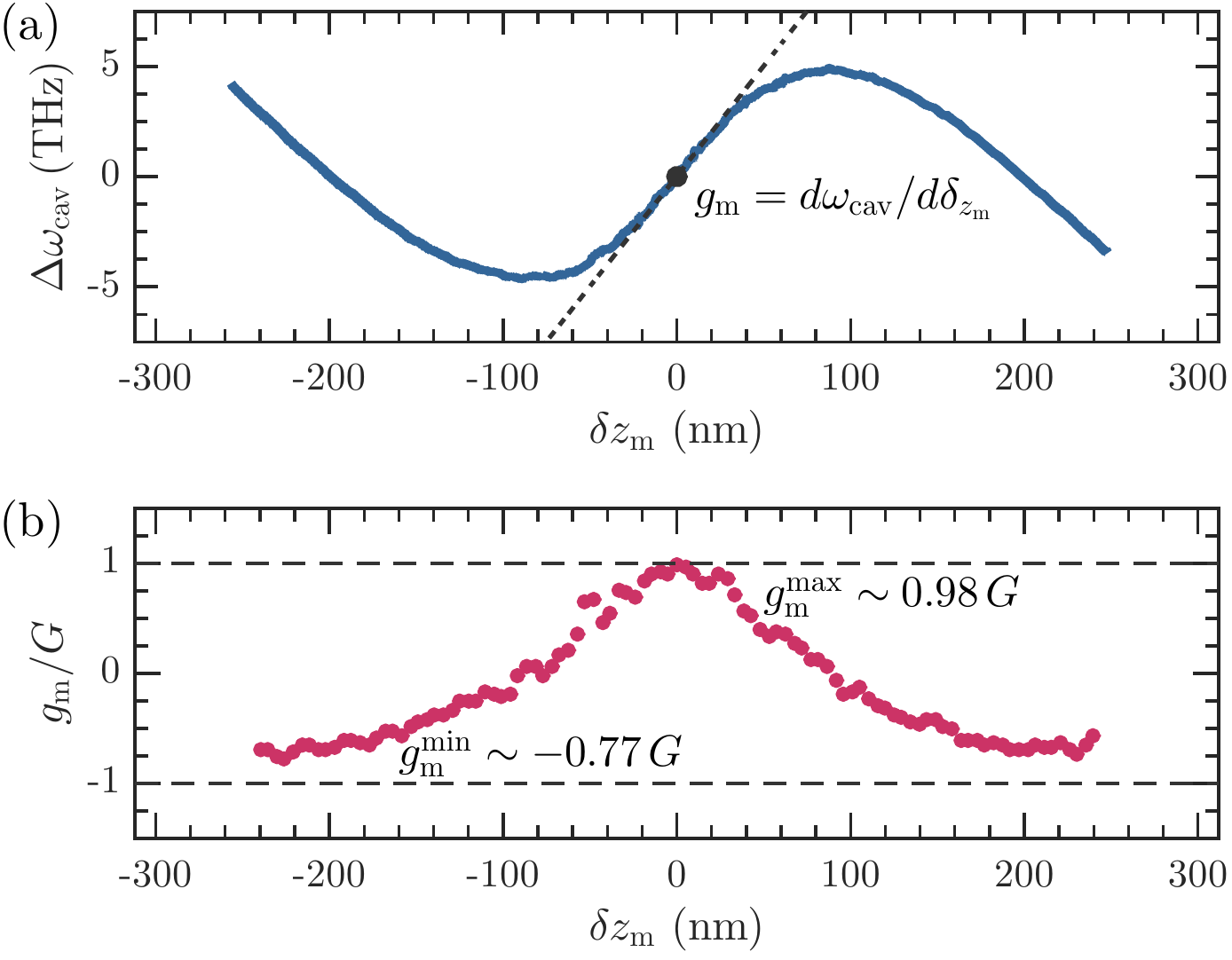}
\caption{\textbf{(a)} Measured detuning of the cavity resonance $\Delta\omega_{\rm cav}$ from its mean value in dependence of the membrane position change $\delta z_{\rm m}$. For this measurement the membrane was near the planar fiber at $L_1\approx5.8$\,$\mu$m. \textbf{(b)} Positional dependence of the normalized optomechanical coupling strength $g_{\rm m}/G$ obtained from the derivative of the data in \textbf{(a)}.} 
\label{Abb:opto} 
\end{figure}

Figure\,\ref{Abb:opto}(a) displays the cavity resonance frequency shift $\delta\omega_{\rm cav}$ when the membrane is moved by $\delta z_{\rm m}$ relative to its rest position $z_{\rm m,0}$ at $L_1\approx5.8$\,$\mu$m. From its derivative displayed in Fig.\,\ref{Abb:opto}(b), we find a maximum optomechanical coupling $g_{\rm m}^{\rm max}=0.98\,G=2\pi\times 15.9$\,GHz/nm. Moreover, $g_{\rm m}\in [g_{\rm m}^{\rm min}, g_{\rm m}^{\rm max}]$ depends on the membrane intracavity position, which shows a slight asymmetry.

\begin{table}
\caption{\label{tab:para} Key parameters of our MiM cavity optomechanical set-up operated at about 480\,mK. All the membrane properties are related to the (1,1)-eigenmode.}
\begin{ruledtabular}
	\begin{tabular}{cccc}
	$\omega_{\rm m}$  		 & $2\pi \times 263.8$\,kHz     & $\omega_{\rm cav}$             & $2\pi \times 384$\,THz \\
	$Q_{\rm m}$            & $10,757,900\pm 1,500$        & $L_{\rm cav}$                  & 23.7\,$\mu$m  \\
	$\gamma_{\rm m}$       & $2\pi \times 24.4$\,mHz      & $\kappa_{\rm cav}$             & $2\pi \times 75$\,GHz     \\
	$m_{\rm eff}$          & $9.7\times10^{-11}$\,kg      & $\mathcal{F_{\rm 0}}$          & 60            \\
	$x_{\rm zpf}$          & $5.7\times10^{-16}$\,m       & $\mathcal{F_{\rm m}}$          & 50 \ldots 120 \\
	$G$           	       & $2\pi \times 16.2$\,GHz/nm 	& $g_{\rm m}/2\pi$               & $(-12.5 \ldots 15.9)$\,GHz/nm\\
	$g_0$  			           & $2\pi \times 9$\,kHz         & $f_{1,1}\times Q_{m} $         & $3\times10^{12}$\,Hz \\
	\end{tabular}
\end{ruledtabular}
\end{table}

To fully characterize our MiM cavity, several cavity optomechanical parameters are of importance. They are listed in Table\,\ref{tab:para}. We define the finesse by $\mathcal{F}=\omega_{\rm FSR}/\kappa_{\rm cav}$, where $\kappa_{\rm cav} = 2\pi\times75$\,GHz is the full width at half maximum of the MiM cavity resonance. In a MiM configuration $\mathcal{F_{\rm m}}$ (like $g_{\rm m}$) depends on the absolute position of the membrane inside the cavity. Since $\omega_{\rm m} \ll \kappa_{\rm cav}$, our MiM cavity is in the so-called bad cavity limit. As mentioned before, this situation cannot be avoided, because coupling a membrane to atoms in an optical lattice via retroreflected light automatically limits the achievable finesse of the cavity.

The zero point fluctuation of the (1,1)-eigenmode is $x_{\rm zpf}=\sqrt{\hbar/2m_{\rm eff}\omega_{\rm m}}=5.7\times10^{-16}$\,m. The maximum single photon optomechanical coupling strength is $g_{\rm 0}=|g_{\rm m}|^{\rm max}\cdot x_{\rm zpf} = 2\pi\times 9$\,kHz. Finally, the product $f_{1,1}\times{Q_{\rm m}} =3 \times 10^{12}$\,Hz of the (1,1)-eigenmode is larger than $k_{\rm B}T_{1,1}/h = 9\times10^{10}$\,Hz. This is an essential condition for quantum ground state cooling using radiation pressure, \citep*{Aspelmeyer2014} which is apparently the case for our set-up. 
  
\section{Summary}

An all-fiber MiM cavity set-up has been constructed in cryogenic UHV environment as well as an apparatus to produce ultra-cold atoms, which can subsequently be loaded into an optical lattice. The MiM cavity can be accurately aligned \textit{in-situ} using two low-temperature compatible fiber positioning stages with five degrees of freedom each. The motional state of the membrane can be detected with a state-of-the-art homodyne detection scheme. At the currently achievable minimal temperature of about 480\,mK, we determined all key parameters that characterize our optomechanical system for a high-stress Si$_{3}$N$_{4}$ membrane. Comparing our values with the requirements suggested in Ref.\,\onlinecite{Vogell2013}, we conclude that it is feasible to realize an atom-membrane HQS with our experimental set-up. This would allow for sympathetically cooling the membrane into its quantum ground state \citep*{Vogell2013, AJ2015} and would moreover pave the way for coherent state transfer and entanglement of the HQS.

\section*{Acknowledgments}
The authors thank M. Nitschke for his help with assembling the all-fiber MiM cavity set-up, O. Hellmig for his help with the fiber processing and our machine shop, particularly R. P. Benecke, J. Path for machining the radiation shields and the cavity housings. Further, we thank the engineering team from Oxford Instruments, particularly H. Agrawal, S. Kingsley and R. Viana, for their technical support. Financial support from the ERC Advanced grant 'FURORE' as well as the DFG via grants SFB668-A5, Wi1277/29-1, BE4793/2-1 and SE717/9-1 is gratefully acknowledged.



\end{document}